\documentclass[aps,prbbib,floatfix,floats]{revtex4}
\usepackage{graphicx,latexsym}
\usepackage{epsfig}
\usepackage{amsmath}
\usepackage{color}
\begin{document}
\title{\bf Theoretical study of electronic transport through a small quantum dot with a magnetic impurity}

\author{Mugurel \c{T}olea}
\affiliation{National Institute of Materials Physics, POBox MG7,
Bucharest-Magurele, Romania} \affiliation{Institute of Molecular
Physics, Polish Academy of Science, ul. M. Smoluchowskiego 17, 60-179 Pozna{\'n}, Poland}
\author{Bogdan R. Bu{\l}ka}
\affiliation{Institute of Molecular Physics, Polish Academy of
Science, ul. M. Smoluchowskiego 17, 60-179 Pozna{\'n}, Poland}

\begin{abstract}
We model a small quantum dot with a magnetic impurity by the
Anderson Hamiltonian with a supplementary exchange interaction term.
The transport calculations are performed by means of the Green
functions within the equation of motion scheme, in which two
decoupling procedures are proposed, for high and low temperatures,
respectively. The paper focuses on the charge fluctuations for such
a system, aspect not addressed before, as well as on the Kondo
resonance. We show a specific role of the excited state, which can
be observed in transport and in spin-spin correlations. Our studies
show on a new many-body feature of the phase shift of transmitted
electrons, which is manifested in a specific dip. In the Kondo
regime, our calculations complement existing theoretical results.
The system shows three Kondo peaks in the density of states: one at
the Fermi energy and two side peaks, at a distance corresponding to
the singlet-triplet level spacing. The existence of the central peak
is conditioned by a degenerate state (the triplet) below the Fermi
energy.
\end{abstract}

\maketitle

\section{Introduction}

Quantum dots (QD) are promising systems for nano-technology because their
quantum-mechanical parameters are easy to control. Motivated, in part, by the
recent developing in the field of magnetic semiconductors \cite{Dietl}, or by
the spintronic advances, theorists and experimentalists have started to address
also the problem of quantum dots with magnetic impurities
\cite{Govorov,Murthy,Aldea,Heersche,Kaul,Rossier}. In such systems, electrons
in the quantum dot are subject to a supplementary exchange interaction. This
induces a competition between the formation of the Kondo cloud and the magnetic
interactions inside QD, but also specific charge fluctuations, aspect less
addressed in literature. Sometimes, the role of the impurity is played by
another quantum dot with odd occupancy \cite{Science}, but also, as in a recent
experiment \cite{Heersche}, the impurity can be a transitional magnetic atom
(Co). The splitting of the Kondo peak was attributed by the authors to the
indirect RKKY interaction \cite{Simon,Vavilov,hoff2006}. We believe that a
central peak would have also been present for a much lower temperature. Govorov
\cite{Govorov} studied theoretically the spectrum and the response to optical
excitations of an InGaAs/GaAs quantum dot with a single Mn impurity. Recently,
Murthy \cite{Murthy} also addressed the problem of interplay between the Stoner
and the Kondo effect in a large QD with a magnetic impurity.

The competition between the Kondo and antiferromagnetic coupling was
also intensively studied in a double-QD geometry coupled by hopping
or by the RKKY interaction (e.g.
\cite{Busser,M,C,Lopez,GeorgesM,Aono,Simon,Vavilov,Silva,hoff2006}).
This is related with another interesting issue: the Kondo effect in
QD with an integer spin \cite{Sasaki,Eto,Giuliano,Izumida,PColeman}.
In the experiment of Sasaki \cite{Sasaki}, the applied magnetic
field can change the ground state and the relative position of the
singlet and triplet states, and it can also induce a degeneracy
point in the energy spectrum. In the vicinity of the degeneracy
point, three Kondo peaks are noticed in the differential
conductance, which is in agreement with our calculations for the
Kondo regime.

Previous theoretical papers have focused mainly on the Kondo effect in large
QDs with magnetic impurities, or on the two-impurity problem. In the present
paper we are also interested in the Kondo resonance, but in a different system:
a small QD with a magnetic impurity that is coupled only to the electrons in
the dot. We are interested in interplay between magnetic interactions: the
dot-leads Kondo coupling (as in the experimental set-up \cite{GG,Crone}) and
the intra-dot magnetic  exchange interactions (between electrons in QD and the
magnetic impurity). We would like to consider spin-flip processes on the
impurity and their detection by the current flowing through the QD, both in
high and low temperature regimes. In spectral terms, the flip of the impurity
spin is associated with singlet-triplet excitations. To our knowledge, the
problem was not considered yet.

Apart from the Kondo regime, we shall also analyze the regime of high
temperatures. The solution in this case is more easy to obtain, and it allows
simple analytical expressions. These will be used to discuss the charge
fluctuations and spin-spin correlations aspects. An electron that enters QD
forms with the magnetic impurity a correlated singlet or a triplet ground
state, respectively to the sign of the exchange parameter. Both situations are
met in experiments \cite{Heersche}. In these cases, the excited states also
play an important role. It can be seen in the conductance peaks, which
amplitude strongly depend on the coupling to the leads, and in the spin-spin
correlation functions, especially when the excited states lie below the Fermi
level.  We also address the problem of the transmission phase. The measurement
of the transmission phase (and not only the amplitude) is possible to realize,
for some time \cite{Yacoby,Schuster}, and can provide interesting supplementary
information. For our system, a dip can be noticed in the phase evolution,
precisely at the excited state position, in condition of constant electronic
occupancy on QD. We will discuss our results in connection with a recent
experiment \cite{Avinun}, where similar features were observed. Our studies are
an explicit demonstration of a generalized Fridel sum rule in the presence of
electronic correlations, the theorem which was just recently proved by Rontani
\cite{Rontani}.

 The paper is organized as follows: in the Section II and III we
describe our model and the Green functions approach. The equation of motion
(EOM) technique is presented in detail. A decoupling scheme adequate for high
temperatures is presented in section IV. The scheme allows to obtain analytical
results for the conductance, dot occupancy, spin-spin correlator and the phase
evolution. In Section V, we get an insight into the Kondo regime using a more
advanced decoupling scheme. One can observe the formation of three Kondo peaks
in the local DOS: at the Fermi energy and two side peaks. The central peak
exists only if a degenerate state (the triplet) is below the Fermi energy.
Section VI contains concluding remarks. For completeness, the spectrum of the
system and a discussion on the formation of the Kondo resonance are presented
in the Appendix A and B, respectively.

\maketitle

\section{Model of a quantum dot with a magnetic impurity}

The considered system of the QD with a magnetic impurity is
presented in Fig. 1. Conduction electrons passing through the QD
interact with accumulated electrons leading to the Kondo resonance
in low temperatures. A magnetic impurity is connected to the
electrons in the QD by the exchange coupling $J$.
\begin{figure}
\centering
 \epsfxsize=0.3\textwidth \epsfbox{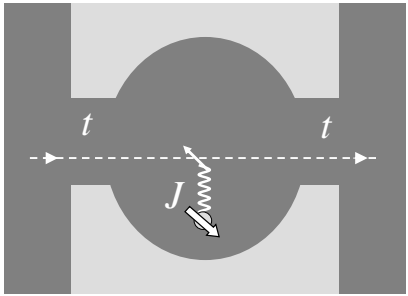}\\
  \caption{Scheme of a quantum dot
coupled to ideal leads through tunnelling barriers $t$. A magnetic impurity
interacts with electrons in the dot through an exchange coupling $J$.}
\end{figure}
The Hamiltonian corresponding to our model is written as
\begin{equation}
H=\sum_{k,\sigma,\alpha}{\epsilon_k
c^\dagger_{k\sigma,\alpha}c_{k\sigma,\alpha}}+\epsilon_0\sum_{\sigma}{c^\dagger_{0\sigma}c_{0\sigma}}+
Uc^\dagger_{0\uparrow}c_{0\uparrow}c^\dagger_{0\downarrow}c_{0\downarrow}+ J\vec{s}\cdot
\vec{S}+\sum_{k,\sigma,\alpha}{t_\alpha(c^\dagger_{0\sigma}c_{k\sigma,\alpha}+h.c.)}\;,
\end{equation}
where the first term represents the electrons in the lead $\alpha= L$, $R$;
the second term stands for the up-most electronic level $\epsilon_0$
in the dot. The the third and the fourth terms describe the
interactions: the Coulomb interaction of electrons with opposite
spin orientation at the level $\epsilon_0$ and exchange interactions
with the magnetic impurity. Here, $\vec{s}$ is the spin operator for
electrons at the level $\epsilon_0$, whereas $\vec{S}$ denotes the
spin operator at the magnetic impurity. It is imposed a single
occupancy at the impurity $d^\dagger_\uparrow d_\uparrow +
d^\dagger_\downarrow d_\downarrow =1$ and the the spin $S=1/2$.
Therefore, $S^z=1/2(d^\dagger_{\uparrow}
d_{\uparrow}-d^\dagger_{\downarrow} d_{\downarrow})$,
$S^+=d^\dagger_{\uparrow}d_{\downarrow}$,
$S^-=d^\dagger_{\downarrow}d_{\uparrow}$, and the exchange term
between the impurity and the spin of the electron localized at the
QD can be written as
\begin{equation}\label{1}
J\vec{s}\cdot\vec{S}=\frac12J(c^\dagger_{0\uparrow} c_{0\uparrow}-c^\dagger_{0\downarrow}
c_{0\downarrow})S^z + \frac12Jc^\dagger_{0\uparrow} c_{0\downarrow} S^- +
\frac12Jc^\dagger_{0\downarrow} c_{0\uparrow} S^+\;.
\end{equation}
The last term in the Hamiltonian (1) corresponds to the coupling
between the quantum dot and the leads. We are interested in
specific correlations effects, so the eventual effects introduced
by an asymmetric coupling are outside our purpose here. It shall
be considered, for simplicity, $t_L=t_R=t$.

The energy spectrum of the isolated QD is easily to calculate (see
the Appendix A). We will show that the ground states, as well as the
excited states, participate in the electronic transport. The
conductance peaks corresponding to the transport through the excites
states are smaller, but they are high enough to be observable in an
experiment.

Our model (1) is a generalization of the single impurity Anderson model (SIAM),
to which a spin-interaction term is added. Recently, Zhang et al. \cite{Zhang}
considered also an extended SIAM model with a a one-body phenomenological
spin-flip term. Their purpose was to study a different situation, namely, the
electronic transport through a quantum dot with spin-orbit interactions coupled
to ferromagnetic leads. In the Hamiltonian (1), the total spin is conserved (in
contrast to the model used in \cite{Zhang}), which is important for the
electronic transport.

{\section{Determination of Green functions in
equation of motion scheme}

Now, we want to determine the conductance and other physical
quantities for the model described by the Hamiltonian (1). The
current can be expressed by means of the non-equilibrium Green
functions as \cite{Meir-Landauer, Haug}
\begin{equation}
j = \frac{ 2e}{h}\Gamma
\int{d\omega[f_L(\omega)-f_R(\omega)](-\mathrm{Im}\langle\langle
c_{0\uparrow}|c_{0\uparrow}^\dagger\rangle\rangle),}
\end{equation}
where $\Gamma=2\pi t^2\rho$ and $\rho=1/2D$ is the DOS for the
square band approximation; the half-width $D$ will be taken as
unity. $f_L$ and $f_R$ are the Fermi distribution functions in the
left and right leads, respectively. $\langle\langle
c_{0\uparrow}|c_{0\uparrow}^\dagger\rangle\rangle$ is the retarded
single particle Green function for an electron with the spin
$\sigma=\uparrow$ at the QD, which can be determined by the equation
of motion (EOM). The differential conductance for the
quasi-equilibrium case can be deduced from the formula (3)
\begin{equation}
{\mathcal G}=\frac{2e^2}{h}\Gamma \int{d \omega (-\frac{\partial
f(\omega)}{\partial\omega}) \mathrm{Im}\langle\langle
c_{0\uparrow}|c_{0\uparrow}^\dagger\rangle\rangle }\;.
\end{equation}

The equation of motion (EOM) for the energy dependent retarded Green function
is given by
\begin{equation}
\omega\langle\langle A|B
\rangle\rangle=\langle\{A,B\}\rangle+\langle\langle
[A,H]|B\rangle\rangle\;,\end{equation} where $\langle\{A,B\}\rangle$
is the thermal average of the anticommutator between the operators A
and B. Whenever necessary, the averages will be calculated by means
of the fluctuation-dissipation theorem $\langle
AB\rangle=-(1/\pi)\int{d\omega f(\omega)\mathrm{Im}\langle\langle
B|A\rangle\rangle}$, where $f(\omega)$ is the distribution function
at the dot. In equilibrium, it is equal with the Fermi distribution
in the leads.

To determine the current (3) one needs the single-particle Green function
$\langle\langle c_{0\sigma}|c_{0\sigma}^\dagger\rangle\rangle$, for which the
EOM is
\begin{eqnarray} (\omega-\epsilon_0)\langle\langle
c_{0\uparrow}|c_{0\uparrow}^\dagger\rangle\rangle&=&1+U\langle\langle
c_{0\uparrow}c_{0\downarrow}^\dagger
c_{0\downarrow}|c_{0\uparrow}^\dagger\rangle\rangle+
\frac12J\langle\langle c_{0\uparrow}S^z|c_{0\uparrow}^\dagger\rangle\rangle\\
&+&\frac12J\langle\langle
c_{0\downarrow}S^-|c_{0\uparrow}^\dagger\rangle\rangle
+2t\sum_k\langle\langle
c_{k\uparrow}|c_{0\uparrow}^\dagger\rangle\rangle\nonumber
\end{eqnarray}
As usual for an interacting problem, one has also to calculate the
many-particle Green functions
\begin{eqnarray}
(\omega-\epsilon_0)\langle\langle
c_{0\uparrow}S^z|c_{0\uparrow}^\dagger\rangle\rangle&=& \langle
S^z\rangle +\frac18J\langle\langle
c_{0\uparrow}|c_{0\uparrow}^\dagger\rangle\rangle+
U\langle\langle c_{0\uparrow}c_{0\downarrow}^\dagger c_{0\downarrow}S^z|c_{0\uparrow}^\dagger\rangle\rangle\\
&-&\frac14J\langle\langle c_{0\downarrow}S^-|c_{0\uparrow}^\dagger\rangle\rangle
+\frac12J\langle\langle c_{0\downarrow}c_{0\uparrow}^\dagger c_{0\uparrow}S^-|c_{0\uparrow}^\dagger\rangle\rangle\nonumber\\
&+&2t\sum_k\langle\langle c_{k\uparrow}S^z|c_{0\uparrow}^\dagger\rangle\rangle\nonumber\\
(\omega-\epsilon_0+\frac14J)\langle\langle
c_{0\downarrow}S^-|c_{0\uparrow}^\dagger\rangle\rangle&=&\frac14J\langle\langle
c_{0\uparrow}|c_{0\uparrow}^\dagger\rangle\rangle+
(U+\frac12J)\langle\langle c_{0\downarrow}c_{0\uparrow}^\dagger c_{0\uparrow}S^-|c_{0\uparrow}^\dagger\rangle\rangle\nonumber\\
&-&\frac12J\langle\langle c_{0\uparrow}S^z|c_{0\uparrow}^\dagger\rangle\rangle
+J\langle\langle c_{0\uparrow}c_{0\downarrow}^\dagger c_{0\downarrow}S^z|c_{0\uparrow}^\dagger\rangle\rangle\nonumber\\
&+&2t\sum_k\langle\langle c_{k\downarrow}S^-|c_{0\uparrow}^\dagger\rangle\rangle\nonumber\\
(\omega-\epsilon_0-U)\langle\langle c_{0\uparrow}c_{0\downarrow}^\dagger
c_{0\downarrow}|c_{0\uparrow}^\dagger\rangle\rangle&=&\langle c_{0\downarrow}^\dagger
c_{0\downarrow}\rangle+
 \frac12J\langle\langle c_{0\uparrow}c_{0\downarrow}^\dagger c_{0\downarrow}S^z|c_{0\uparrow}^\dagger\rangle\rangle+
\frac12J\langle\langle c_{0\downarrow}c_{0\uparrow}^\dagger c_{0\uparrow}S^-|c_{0\uparrow}^\dagger\rangle\rangle\nonumber\\
&+&2t\sum_k\langle\langle c_{k\uparrow}c_{0\downarrow}^\dagger
c_{0\downarrow}|c_{0\uparrow}^\dagger\rangle\rangle
-2t\sum_k\langle\langle c_{0\uparrow}c_{k\downarrow}^\dagger c_{0\downarrow}|c_{0\uparrow}^\dagger\rangle\rangle\nonumber\\
&+&2t\sum_k\langle\langle c_{0\uparrow}c_{0\downarrow}^\dagger
c_{k\downarrow}|c_{0\uparrow}^\dagger\rangle\rangle\nonumber
\end{eqnarray}
and
\begin{eqnarray}
(\omega-\epsilon_0-U)\langle\langle c_{0\uparrow}c_{0\downarrow}^\dagger
c_{0\downarrow}S^z|c_{0\uparrow}^\dagger\rangle\rangle&=& \langle c_{0\downarrow}^\dagger
c_{0\downarrow}S^z\rangle+\frac18J\langle\langle c_{0\uparrow}c_{0\downarrow}^\dagger
c_{0\downarrow}|c_{0\uparrow}^\dagger\rangle\rangle
+\frac14J\langle\langle c_{0\downarrow}c_{0\uparrow}^\dagger c_{0\uparrow}S^-|c_{0\uparrow}^\dagger\rangle\rangle\\
&+&2t\sum_k\langle\langle c_{k\uparrow}c_{0\downarrow}^\dagger
c_{0\downarrow}S^z|c_{0\uparrow}^\dagger\rangle\rangle
-2t\sum_k\langle\langle c_{0\uparrow}c_{k\downarrow}^\dagger c_{0\downarrow}S^z|c_{0\uparrow}^\dagger\rangle\rangle\nonumber\\
&+&2t\sum_k\langle\langle c_{0\uparrow}c_{0\downarrow}^\dagger c_{k\downarrow}S^z|c_{0\uparrow}^\dagger\rangle\rangle\nonumber\\
(\omega-\epsilon_0-\frac14J-U)\langle\langle c_{0\downarrow}c_{0\uparrow}^\dagger
c_{0\uparrow}S^-|c_{0\uparrow}^\dagger\rangle\rangle&=& -\langle c_{0\uparrow}^\dagger
c_{0\downarrow}S^-\rangle+\frac14J\langle\langle c_{0\uparrow}c_{0\downarrow}^\dagger
c_{0\downarrow}|c_{0\uparrow}^\dagger\rangle\rangle
+\frac12J\langle\langle c_{0\uparrow}c_{0\downarrow}^\dagger c_{0\downarrow}S^z|c_{0\uparrow}^\dagger\rangle\rangle\nonumber\\
&+&2t\sum_k\langle\langle c_{k\downarrow}c_{0\uparrow}^\dagger
c_{0\uparrow}S^-|c_{0\uparrow}^\dagger\rangle\rangle
-2t\sum_k\langle\langle c_{0\downarrow}c_{k\uparrow}^\dagger c_{0\uparrow}S^-|c_{0\uparrow}^\dagger\rangle\rangle\nonumber\\
&+&2t\sum_k\langle\langle c_{0\downarrow}c_{0\uparrow}^\dagger
c_{k\uparrow}S^-|c_{0\uparrow}^\dagger\rangle\rangle\nonumber
\end{eqnarray}
In the absence of a magnetic field, one can consider $\langle
S^z\rangle=0$, $\langle n_{0\downarrow}\rangle=\langle
n_{0\uparrow}\rangle=\langle c_{0\uparrow}^\dagger
c_{0\uparrow}\rangle$ and $\langle c_{0\downarrow}^\dagger
c_{0\downarrow}S^z\rangle=-\frac12\langle c_{0\uparrow}^\dagger
c_{0\downarrow}S^-\rangle=-\frac13\langle
\vec{s}\cdot\vec{S}\rangle$, $(S^z)^2=\frac14$ (see also
\cite{Nagaoka} for some details on how the constriction
$d^\dagger_\uparrow d_\uparrow + d^\dagger_\downarrow d_\downarrow
=1$ is used).

We have got a set of six equations, Eqs.(6)-(8), for all Green functions at the
quantum dot. These functions are coupled with other many-particle Green
functions $\langle\langle c_{k\sigma} A |c_{0\uparrow}^\dagger\rangle\rangle$,
which describes the coupling between QD and the electrodes. The EOM procedure
requires to write also the equation of motion for the new set of Green
functions, which will be coupled with another new set of the higher order
many-particle Green functions. In order to close the infinite series of the
equations of motion, one has to decouple many-particle Green functions. The
simplest approximation is
\begin{equation}\label{HT}
2t\sum_k\langle\langle
c_{k,\sigma}A|c^\dagger_{0,\sigma}\rangle\rangle\approx -i\Gamma
\langle\langle c_{0,\sigma}A|c^\dagger_{0,\sigma}\rangle\rangle\;.
\end{equation}
 The approximation takes into account charge
fluctuations between the electrode and QD, but it neglects the
spin-flip processes and the Kondo resonance. This approximation
correctly describes the limit of the Coulomb blockade and gives
the same results as those obtained by Nagaoka \cite{Nagaoka} and
by Meir {\it et al} \cite{Meir91} (for QD without impurity) in
high temperatures. It is important to notice that all Green
functions at QD are treated exactly, so the intra-dot
correlations are fully accounted. This approximation will be used
in the next section to study the high temperature properties of
our model.

\section{Studies of the high temperatures regime}

Using the approximation (\ref{HT}) we get a set of six linear
equations, with the correlators $\langle n_{0\uparrow}\rangle$ and
$\langle c_{0\uparrow}^\dagger c_{0\uparrow}S^z\rangle$ as
parameters. One can easily find analytical expressions for all
many-particle Green functions at the QD. Next, the correlators are
calculated by means of the appropriate Green functions $\langle
n_{0\uparrow}\rangle=-(1/\pi) \int{d\omega
f(\omega)\mathrm{Im}\langle\langle c_{0\uparrow}|
c_{0\uparrow}^\dagger\rangle\rangle}$ and $\langle
c_{0\uparrow}^\dagger c_{0\uparrow}S^z\rangle=-(1/\pi)
\int{d\omega f(\omega)\mathrm{Im}\langle\langle c_{0\uparrow}S^z|
c_{0\uparrow}^\dagger\rangle\rangle}$. In order to calculate these
integrals one has first to decompose the Green functions in simple
fractions. The solutions are expressed by the digamma function
\cite{Hamman}, which at zero temperature, turns into the arctan
function.

The main results of this section are presented in Fig.2. The upper
panels show the conductance (solid curves) and the electron
occupancy  of the QD (dotted curves) as a function of the position
of the dot energy $\epsilon_0$ for different parameters $\Gamma$.
The lower panels show the corresponding behavior of the spin-spin
correlator.

For $J>0$ -- column a) -- four conductance peaks are clearly seen,
while for $J<0$ -- column b) -- there are two pronounced peaks and
two much smaller peaks. To explain the origin of the peaks, let us
start with a high value of $\epsilon_0$, which is much above the
Fermi energy $E_F=0$. For this case the dot is empty.
When $\epsilon_0$ is lowered (by an applied gate
potential), a first electron is allowed to jump into the QD on the
lowest available level. This is the singlet state, for case a) -see Appendix A- and
it is located at $\epsilon_0=0.15$. For case b), the lowest state
is the triplet state at $\epsilon_0=0.05$. By further lowering of
$\epsilon_0$, the excited state energy becomes aligned with the
Fermi level. It is the triplet state at $\epsilon_0=-0.05$, or the
singlet state at $\epsilon_0=-0.15$, for the case a) and b)
respectively. At these positions the conductance shows peaks, for
which the height is smaller than for those corresponding to the
ground states. We do not observe any changes of the electron
occupancy at these positions.

The transmission through the excited states, where the total
occupancy on the dot is constant, can be explained by the
overlapping with the ground state and fluctuations of the charge
distribution between levels. The electron has a probability to
exist on the excited state, and from here it can tunnel into the
leads. This explains the strong coupling-dependence of the excited
states conductance. Notice a significant increase of the peaks
corresponding to the transport through excited states as the coupling $\Gamma$
increases from $0.01$ to $0.08$. Experimentalists indeed have measured such
small conductance peaks corresponding to transport through excited
states (e.g. \cite{Rogge, Sigrist}), when the total occupancy on
the QD is constant. Apart from describing an impurity dot, our model may
be one of the simplest to allow the analysis of excited states behavior
in transport.

The transmittance through the excited states is
more pronounced in Fig2.a than in Fig.2b. This is because the triplet state
(that is the excited state in the case a) is three-fold degenerate and in
favored in transport in comparison with the non-degenerate singlet state. From
the same reason, one can notice that, on the contrary, the peak corresponding
to the ground state (triplet) in Fig.2b is higher than in Fig.2a (singlet). The
conductance peaks corresponding to the transport through the ground states
remain then practically unchanged, only a broadening effect is noticed. A
detailed discussion of the peaks amplitude will be given further in this
section.

 We remind that the resonances in conductance appear when two quantum
states with consecutive occupancy are energetically available in the
same time, allowing for charge fluctuations. So, one can find the
positions of the resonances with the equation $E(\nu,n)=E(\nu\,',n\pm1)$, meaning
that an energy from the spectrum with $n$ electrons becomes equal
with an energy for $n\pm 1$ electrons - see Appendix A. By further
lowering of $\epsilon_0$, the energy for double occupancy will
become equal with the excited single-particle energy. This will
generate a small peak in conductance, and no variation of occupancy.
Finally, a big peak is noticed corresponding to the adding of the
second electron. Actually, a particle-whole symmetry is easy to
notice in Fig.2, and we shall focus the further discussion on the
first two peaks from the right (marked by the vertical, dotted
lines).

The lower panel in Fig.2 shows the evolution of the spin-spin
correlator $\langle\vec{s}\cdot\vec{S}\rangle$, which is a measure
of the coupling between the spin of an electron accumulated at the
QD and the magnetic impurity spin. For $J>0$, electrons at the QD
are antiferromagnetic  coupled with the impurity spin, and
therefore, the spin-spin correlator is negative. Its minimal value
is $-0.7$, $-0.5$ and $-0.3$ for $\Gamma=0.01$, $0.04$ and $0.08$,
respectively. For the case $\Gamma=0.01$, the length of the total
spin, $\langle \vec{S}^2_{tot}\rangle=\langle
\vec{s}^2\rangle+2\langle \vec {s}\cdot
\vec{S}\rangle+\langle\vec{S}^2\rangle$ is then reduced to $0.1$,
i.e. it is almost compensated. When the triplet state gets below
Fermi energy, the correlator $\langle\vec{s}\cdot\vec{S}\rangle$
increases, because the triplet state favors ferromagnetic
coupling. Its maximal value is $<\vec{s}\cdot \vec{S}>\approx
-0.1$, which means $<\vec{S}^2_{tot}>\approx 1.3$ close to $3/2$ -
the value of the magnetic moment for two free $S=1/2$ spins - see
the central part of the graph. The result is important because it
shows that the excited states contribute to the formation of the
local magnetic moment at the QD. It can have also influence on the
Kondo singlet formation with the electrons from the leads. Section
V will show that indeed this is the case.

The evolution of the spin-spin correlator for the ferromagnetic case, presented
in Fig.2b, can be explained in the same way. When the triplet state crosses
$E_F$ the spin-spin correlator reaches maximal values $0.23$, $0.2$ and $0.15$,
for $\Gamma=0.01$, $0.04$ and $0.08$, respectively. The total magnetic moment
increases up to $1.94$, $1.86$ and $1.7$. For each of the three situations, the
spin-spin correlator and the total moment begin to decrease when the singlet
state goes below $E_F$.

\begin{figure}[ht]
\centering
 \epsfxsize=0.9\textwidth \epsfbox{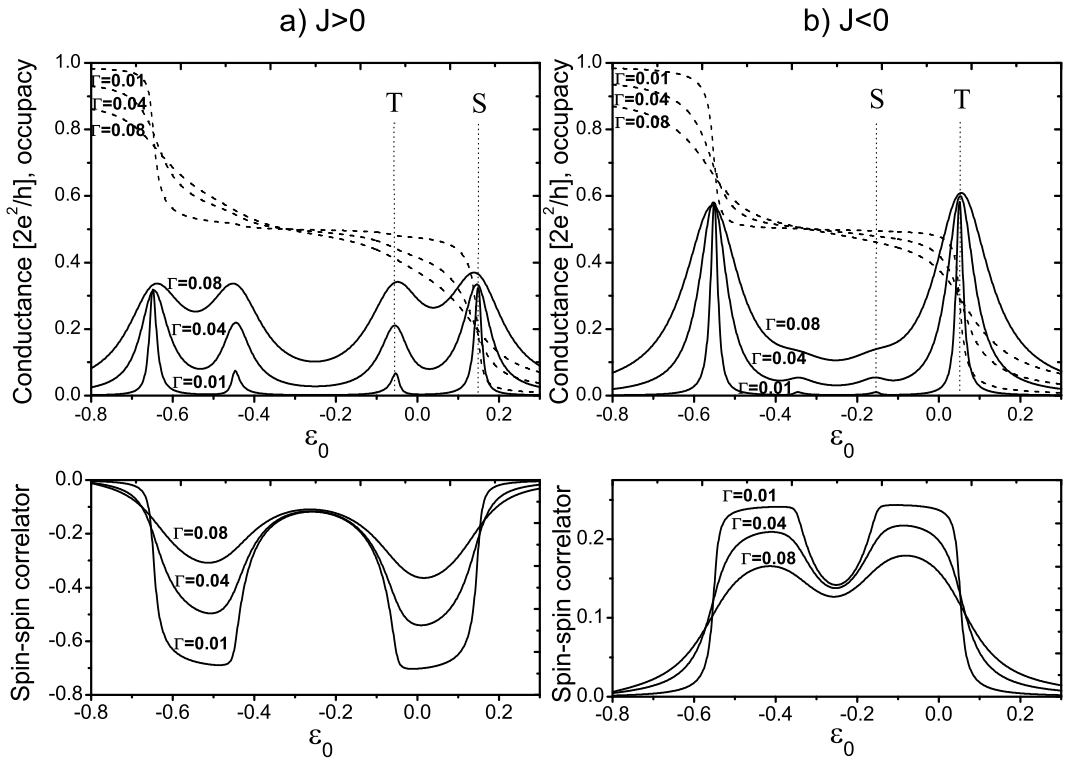}
  \caption {The conductance (solid curves), the dot occupancy per
  spin $\langle n_{0\sigma}\rangle$ (dotted curves) -- upper panels,
  and the spin-spin correlator  $\langle\vec{s}\cdot\vec{S}\rangle$ -- lower panels,
  plotted versus the dot energy $\epsilon_0$, for $U=0.5$, $|J|=0.2$, $E_F=0$, $T=10^{-3}$,
  and for different values of $\Gamma= 0.01$, $0.04$ and $0.08$.
  The vertical lines indicate the position of the Singlet and the Triplet levels.}
\end{figure}

We want to discuss in more details the heights of the conductance peaks
corresponding to the singlet and the triplet states. A simple analytical
solution can be obtained for the conductance, for $T=0$ and in the large $U$ limit
, for the singlet and triplet resonances
\begin{equation}
{\mathcal G}={\mathcal G}_S+{\mathcal G}_T
=\frac{1+2\phi_T}{2(3+4\phi_S-4\phi_S\phi_T)}R_S+\frac{3(1+2\phi_S)}
{2(3+4\phi_S-4\phi_S\phi_T)}R_T\;,
\end{equation}
where
$R_S=(2e^2/h)\times\Gamma^2/[\Gamma^2+(E_F-\epsilon_0+3J/4)^2]$ and
$R_T= (2e^2/h)\times\Gamma^2/[\Gamma^2+(E_F-\epsilon_0-J/4)^2] $
corresponds to the resonant conductance through the singlet and the
triplet level, respectively,
$\phi_S=\arctan[(\epsilon_0-E_F-3J/4)/\Gamma]/\pi$ and
$\phi_T=\arctan[(\epsilon_0-E_F+J/4)/\Gamma]/\pi$. The fractions
that multiply the resonances $R_{S(T)}$ contain the information
about the height of the conductance peaks. The ground and excited
states role changes with the sign of $J$, and the dependence on the
coupling also changes, being much more pronounced for the excited
state. All these effects are incorporated in formula (10). This
zero-temperature formula is given for its simplicity, but it
describes well also situations like in Fig.2, where temperature
$T=10^{-3}$ is considerably lower than the broadening due to
coupling with the leads ($\Gamma=0.01$ to $0.08$). For higher
temperatures, one has to replace in formula (10):
$\phi_S=-\mathrm{Im}\{\Psi[1/2-i(\epsilon_0-E_F-3J/4+i\Gamma)/(2\pi
T)]\}/\pi$, $\phi_T$ in a similar way ($\Psi$ is the digamma
function, see also \cite{Bulka2004}), and then perform an integral required by the conductance
formula deduced from Eq.3. Fig.2 was plotted in this way, with the
complete formula.

By employing further the simplified form Eq.(10), one can estimate the ratio between
the heights of the conductance for the singlet and triplet peaks.
We notice that in Eq.(10), the parameter $\phi_S$ ($\phi_T$) vanishes at the singlet
(triplet) resonant level, and therefore, one can straightforward
calculate the ratio of the conductance peaks
\begin{equation}
\frac{\max[\mathcal {G}_{S}]}{\max[\mathcal{G}_{T}]}=\frac{(1+2\phi_0)(3-4\phi_0)} {9(1-2\phi_0)}\,,
\end{equation}
where $\phi_0=\arctan(J/\Gamma)/\pi$. For the antiferromagnetic
coupling the electronic transmission through the singlet state
dominates and in the limit  $J\gg \Gamma$ this ratio goes to
infinity and the triplet peak disappears. In the case of the
ferromagnetic coupling and $|J|\gg \Gamma$ we have opposite
situation - the singlet states disappears from transport and only
the peak corresponding to the triplet ground state is visible.

In \cite{Aldea,Ramsak}, the authors model a similar spin-dependent
transport situation (for the QD and QPC, respectively), but the
correlations and the effects of the Fermi sea are neglected. Their
problem is analogous to the electron collision with a hydrogen
atom (see \cite{landau}). In such a case, by averaging all
possible scattering processes, the height of the singlet and
triplet peaks will be proportional with their degeneracy resulting
the $1/3$ peak height ratio - independent of the sign of $J$, also
independent of the coupling to the leads. Formulas (10)-(11) are a natural
generalization. The difference results from the charge
accumulation and the Coulomb blockade effect.

In the final part of this section, we analyze the evolution of
the phase shift for the scattered electrons. The phase problem has
raised considerable interest in mesoscopic physics \cite{Hack}. The
principle of the phase shift experimental measurement
\cite{Yacoby,Schuster} uses an interference process, which has been
realized by inserting QD - for which the phase is measured - in
one arm of an Aharonov-Bohm interferometer.
The electronic wave passing through QD acquires a phase shift and
then interferes with the wave travelling through the reference arm.
In this way, one can
measure not only the amplitude of the transmission, but also its
phase, which is extracted from the phase of the Aharonov-Bohm
conductance oscillations.
The authors expected a phase evolution in concordance with the Fridel sum rule,
growing with $\pi$ on resonances, and with constant values between them.
Instead, an intriguing universal phase-lapse effect
was noticed \cite{Yacoby,Schuster}, meaning that the phase has a
drop of $\pi$ between any pair of resonances. The aspect has not yet
received a satisfactory explanation. Recently, Avinun-Kalish et al. \cite{Avinun} realized
a similar experiment, but with a supplementary possibility to control the exact number of
electrons in QD (up to 20 electrons). For the first few electrons (about $10$),
the phase shift had a complex
non-universal behavior, with dips rather than phase-lapses, and for the next electrons,
the "universal" behavior was recovered. Rontani
demonstrated, in a very recent paper \cite{Rontani}, that such features in the phase
evolution are normal to happen, as the mesoscopic systems with electronic correlations do
not obey the classical Fridel sum rule. It means that the
experimentally measured phase shift is not equal to $\pi$ multiplied
with the accumulated charge, but may have variations in the case of
electronic correlations in the system. We explicitly prove the
existence of such dips in the phase evolution, for the model discussed
 in this paper. Moreover, they receive a spectral
interpretation, as being a fingerprint of the excited states.

For simplicity, we consider here only the situation $\Gamma=0.01$,
when the singlet and triplet peaks are well-separated (but the
conclusions are general). One could naively expect that the phase
grows with $2\pi$ on the two separate peaks, or that the phase
evolution obeys the Fridel sum rule and accurately follows the
occupancy curves. However, both hypotheses are wrong, as the
ground state is decisively influencing the phase evolution on the
excited state. We must remember that the very existence of the
excited states peaks is attributed to the overlapping with the
ground states peaks.

The phase shift can be expressed as the argument of the
single-electron Green function \cite{Levi-Buttiker, Aldea}
\begin{equation}
\Phi = \arctan\left[\frac{\mathrm{Im}\langle\langle
c_{0\sigma}|c_{0\sigma}^\dagger\rangle\rangle_{\omega=E_F}}{\mathrm{Re}\langle\langle
c_{0\sigma}|
c_{0\sigma}^\dagger\rangle\rangle_{\omega=E_F}}\right]\;.
\end{equation}
In a single-particle scattering picture, the phase defined as (12)
is equal to the phase-shift of the electron wave-function. The
definition is identical with the definition of
$\theta_t$ in \cite{Levi-Buttiker}. The authors
define the phases as the argument of the scattering matrix elements,
which are the retarded Green functions.

\begin{figure}[ht]
\centering
 \epsfxsize=0.35\textwidth \epsfbox{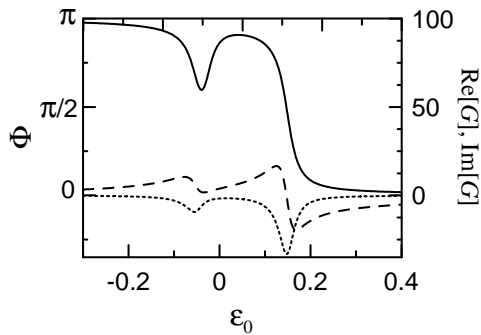}\\
  \caption{Evolution of the phase shift $\Phi$ (solid curve), the real and the imaginary part of the single particle Green function $G=\langle\langle c_{0\sigma}|c_{0\sigma}^\dagger\rangle\rangle_{\omega=E_F}$
  (dashed and dotted curves - in 1/D units) for the same parameters as in Fig.2a and $\Gamma=0.01$.}
\end{figure}

 Fig.3 presents the
phase evolution, the real and the imaginary part of the Green
function. The phase evolves with $\pi$ on the peak corresponding to
the ground state (around $\epsilon_0=0.15$), where an electron is
added. Then the phase presents a dip on the excited state peak
(around $\epsilon_0=-0.05$), in conditions of unchanged dot
occupancy (see Fig.2). Notice that the real part of the Green
function (the dashed curve in Fig.3) changes sign at the ground state
position, but not at the excited state position, which is why the
argument of the complex Green function (12) remains in
the same quadrant in the latter case. The phase
exactly at the triplet position is easy to calculate. We obtained
the minimal value in the dip $(\pi - \arctan[6/\pi])\simeq 0.65\pi$
\cite{arctan}, for a low coupling (see Fig.3). For the negative
exchange parameter (as in Fig.2, column b), and the singlet excited
state, the minimal value for the dip is $(\pi -
\arctan[2/(3\pi)])\simeq 0.93\pi$.

The observed dip is a signature of the excited states and can be
used to probe the existence of such states through phase
measurements. Also, phase measurements can, in principle, provide information
about the nature of the excited states - we obtained the amplitude of the dip
different for the singlet and triplet.

Our phase calculations can be correlated with the
recent experiment by Avinun et al. \cite{Avinun}, where such dips in the phase
evolution were observed. As mentioned, the authors measured the transmission
phase, and have reported a non-universal behavior. The measured
phase showed the expected evolution with $\pi$ on the resonances,
but the authors also observed a number of dips of variable amplitude
(but less than $\pi$).
 Rontani \cite{Rontani} proved that
the phase shift is not necessarily an integer of $\pi$ between two
consecutive Coulomb blockade regions, if a complete formula is
employed. The author stated that values lower than $\pi$ for the
phase, through regions with constant occupancy, are a fingerprint of
electronic correlations. The dips may even reach the value zero for
the case of the spin blockade.

\section{An insight on the Kondo regime}

In this section, we analyze the Kondo physics for our system, in a
low temperature. In this regime, the electron from the dot may
develop a Kondo-correlated state with the electrons in the leads.
This is in competition with the spin-coupling with the magnetic
impurity.

A study of the Kondo resonance requires a method which takes into account
correlations with electrons in the leads and the virtual spin-flip processes.
One has to use the higher order decoupling scheme than that one used in the
previous section. The next step of EOM involves the Green functions with one
operator from the leads and decoupling for the Green functions with two
operators from the leads. Here, an example of EOM
\begin{eqnarray}
(\omega-\epsilon_k)\langle\langle c_{0\uparrow}c_{0\downarrow}^\dagger
c_{k\downarrow}S^z|c_{0\uparrow}^\dagger\rangle\rangle =\langle c_{0\downarrow}^\dagger
c_{k\downarrow}S^z\rangle+t\langle\langle c_{0\uparrow}c_{0\downarrow}^\dagger
c_{0\downarrow}S^z|c_{0\uparrow}^\dagger\rangle\rangle
+2t\sum_q\langle\langle c_{q\uparrow}c_{0\downarrow}^\dagger c_{k\downarrow}S^z|c_{0\uparrow}^\dagger\rangle\rangle\nonumber\\
-2t\sum_q\langle\langle c_{0\uparrow}c_{q\downarrow}^\dagger
c_{k\downarrow}S^z|c_{0\uparrow}^\dagger\rangle\rangle
+\frac14J\langle\langle
 c_{0\uparrow}c_{0\downarrow}^\dagger c_{k\downarrow}|c_{0\uparrow}^\dagger\rangle\rangle
+\frac14J\langle\langle c_{k\downarrow}c_{0\downarrow}^\dagger c_{0\downarrow}S^-|c_{0\uparrow}^\dagger\rangle\rangle\nonumber\\
+\frac14J\langle\langle c_{k\downarrow}c_{0\uparrow}^\dagger
c_{0\uparrow}S^-|c_{0\uparrow}^\dagger\rangle\rangle -\frac12J\langle\langle
c_{k\downarrow}c_{0\uparrow}^\dagger c_{0\uparrow}c_{0\downarrow}^\dagger c_{0\downarrow}
S^-|c_{0\uparrow}^\dagger\rangle\rangle\;.
\end{eqnarray}
The higher order Green functions are decoupled according the following scheme
\begin{eqnarray}
\langle\langle c_{0\uparrow}c_{q\downarrow}^\dagger
c_{k\downarrow}|c_{0\uparrow}^\dagger\rangle\rangle &\approx &\langle c_{q\downarrow}^\dagger
c_{k\downarrow}\rangle \langle\langle c_{0\uparrow}|c_{0\uparrow}^\dagger\rangle\rangle \approx
\delta_{kq}f(\epsilon_k)
\langle\langle c_{0\uparrow}|c_{0\uparrow}^\dagger\rangle\rangle\;,\\
\langle\langle c_{0\uparrow}c_{q\uparrow}^\dagger
c_{k\downarrow}S^-|c_{0\uparrow}^\dagger\rangle\rangle&\approx& \langle c_{q\uparrow}^\dagger
c_{k\downarrow}S^-\rangle \langle\langle c_{0\uparrow}|c_{0\uparrow}^\dagger\rangle\rangle\approx
\delta_{kq}\lambda f(\epsilon_k)\langle c_{0\uparrow}^\dagger c_{0\downarrow}S^-\rangle
\langle\langle c_{0\uparrow}|c_{0\uparrow}^\dagger\rangle\rangle\nonumber\;,\\
\langle\langle c_{0\uparrow}c_{q\downarrow}^\dagger c_{k\downarrow}
S^z|c_{0\uparrow}^\dagger\rangle\rangle&\approx &\delta_{kq}f(\epsilon_k)\langle\langle
c_{0\uparrow}S^z|c_{0\uparrow}^\dagger\rangle\rangle +\delta_{kq}\lambda f(\epsilon_k)\langle
c_{0\downarrow}^\dagger c_{0\downarrow}S^z\rangle\langle\langle
c_{0\uparrow}|c_{0\uparrow}^\dagger\rangle\rangle\;.\nonumber
\end{eqnarray}
The decoupling procedure follows that one proposed by
Meir\cite{Meir91} (also used e.g. in \cite{Meir93,Zhang,Guo}) and by
Nagaoka \cite{Nagaoka} for the Green functions that also contain the
localized spin operator. Only the averages that conserve the total
spin are kept. The system of equations must be closed, so we need to
express further the spin-spin correlator $\langle
c_{q\uparrow}^\dagger c_{k\downarrow}S^-\rangle$ between the
electrons in the leads and the localized spin. This is considered to
be proportional to the dot spin-spin correlator, by a parameter
$\lambda<1$. The choice is justified by the fact that the localized
spin is only indirectly coupled to the electrons in the leads. We
have checked that the main results are not qualitatively influence
for any $0<\lambda <1$. In the numerical calculations we put
$\lambda=t^2/D^2$.

After the decoupling scheme (14), the set of Eqs.(13) (there are 14
equations) becomes linear and is easy to solve, in the
$U\rightarrow\infty$ limit. The solution for the Green functions
contains the fractions: $1/(\omega-\epsilon_k)$,
$1/(\omega-\epsilon_k+J)$ and $1/(\omega-\epsilon_k-J)$. Next, we
come back to the initial system Eqs.(6)-(8), where the summation over
$k$ leads to the appearance of three digamma functions \cite{Hamman}
\begin{equation}
F_a({\omega})\equiv
-\sum_k{\frac{f(\epsilon_k)}{\omega-\epsilon_k+a}}=i\frac\pi{2}+\ln\left(\frac{2\pi
T}{D}\right)+\Psi\left(\frac12-i\frac{\omega+a}{2\pi T}\right)
\end{equation}
with $a=0$ and $\pm J$. Here, $\Psi$ denotes the digamma function.
For $T<T_K$, the function $F_a$ has a pronounced maximum around
$\omega=-a$.

Fig.4 presents the density of states (DOS) of our system for three different
cases: a) when both levels, singlet and triplet, are below $E_F$, b) the
triplet state is below $E_F$, but the singlet state is above, and c) the
singlet state is below $E_F$ and the triplet above $E_F$. In the panel a), the
singlet and triplet peaks overlap, due to the small $J/\Gamma$ ratio. Our model
gives three peaks in the DOS: one at the Fermi level $E_F=0$ and two
symmetrical side peaks, at $\pm J$, corresponding to the basic excitations in
the system. The central peak can be associated with the three-fold degeneracy
of the triplet state. We point out that in the measurements of Sasaki (see
Fig.4a in \cite{Sasaki}) such a three-peak structure can be seen in the case of
the triplet ground state as well as for the singlet ground state, where the
"zero" peak is however much less pronounced and the other two peaks are
considerably enhanced.

For a large level spacing (large $J$), only the peak at $E_F$ is of interest.
The distant side peaks would not influence the conductance in any way. Our
calculations show that the central Kondo peak is formed when the triplet state
is below $E_F$ (Fig4.b), and it is not formed when only the singlet state is
below $E_F$ (Fig4.c). Such a result was expected because the singlet state is
not degenerate, and a Kondo peak at the Fermi energy can only appear when the
system has a degenerate occupied level (it is not exclusively a spin effect,
but, more generally, it originates in the Pauli exclusion principle).

\begin{figure}[ht]
\centering
 \epsfxsize=0.35\textwidth \epsfbox{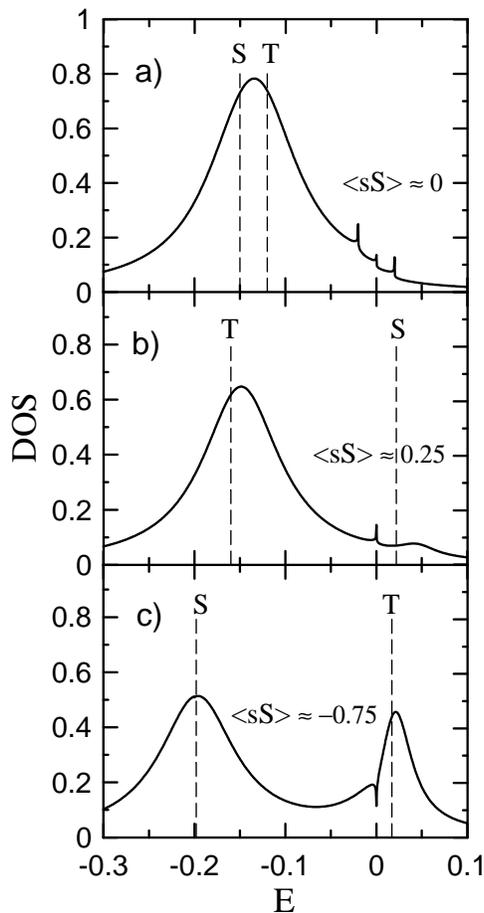}\\
  \caption{Plots of the local density of states (DOS) for the Kondo regime:
  a) $\epsilon_0=-0.15$ and $J=0.02$;
  b) $\epsilon_0=-0.11$ and $~J=-0.2$; c) $\epsilon_0=-0.05$ and $J=0.2$.
  The corresponding value of the spin-spin correlator
  is written on each graph.
  The vertical lines indicate the positions of the singlet and triplet
  levels. For all the three plots $E_F=0$, $\Gamma=0.025$ and $T=10^{-6}$.} \end{figure}

In Appendix B we show that the existence of the central Kondo peak can be
related with the value of the spin-spin correlator. The peak is absent for a
strong antiferromagnetic coupling of the spins inside the dot. Such a case
happens when the singlet state is below the Fermi energy and the triplet above.
Therefore, the arguments based on the value of the spin-spin correlator (given
in Appendix B) lead to the same conclusion as the arguments based on degeneracy
considerations (presented in the previous paragraph).

In this section we have focused exclusively on the description of
the Kondo peaks in the DOS. However, the differential conductance
accurately follows the DOS. In \cite{K}, the authors prove the
decoupling of Meir \cite{Meir91} to give physically correct results
(qualitatively) in the Kondo regime, but the decoupling cannot be
used in the mixed-valance regime - when charge fluctuations are
mixed with spin fluctuations. This is why, in this section, we
restricted the analysis to the spin fluctuations and the Kondo
regime, when the occupancy on the dot is $\approx 1$. For the
mixed-valance regime, the approximation made by Lacroix
\cite{Lacroix} for the SIAM should be used instead. A generalization
of \cite{Lacroix} was only done for systems with a complex
noninteracting network around the impurity
\cite{Bulka2001,Hofs,Bulka2004PRL,Ora,K,Stefanski}. In the present
paper, we added an interaction term, considerably increasing the
complexity, and believe that the scheme \cite{Lacroix} may no longer
be applicable. The more simple decoupling used in section IV allows
the analysis of charge fluctuations, but looses the spin
fluctuations - so the two decouplings presented in this paper are
complementary.

 \section{Conclusions}

The paper addresses the problem of electronic transport through a
small quantum dot with a magnetic impurity. Our Hamiltonian model
is the single impurity Anderson extended by a term of an exchange
interaction between the electrons at QD and the magnetic impurity.
The calculations have been performed
by means of the Green functions method within the EOM scheme and presented in detail,
with two different decoupling proposed, in high and low temperatures,
respectively.

We have studied the charge fluctuations for high temperatures, the aspect not
addressed before for the case of QDs with magnetic impurities. The singlet and
triplet conductance peaks are shown to depend on the ratio
$J/\Gamma$ (the exchange interaction versus the coupling strength with the
leads). In particular, the transmission through the excited states depends
strongly on the coupling to the leads, while the
transmission through the ground states show just a broadening effect.

The spin-spin coupling $\langle\vec{s}\cdot\vec{S}\rangle$ between
the magnetic impurity and the electrons in the dot is also analyzed.
The electron that enters the dot forms with the magnetic impurity a
correlated singlet or a triplet ground state, respectively to the
sign of the exchange parameter. The absolute value of the correlator
$|\langle\vec{s}\cdot\vec{S}\rangle|$ reaches then the maximal
value. When the excited states are shifted below the Fermi energy,
the spin-spin coupling decreases. In the regime when the levels are
dip below $E_F$, the spins behave like free spins and
$|\langle\vec{s}\cdot\vec{S}\rangle|$ reaches its minimal value.
This is due to an internal charge redistribution between the ground
and the excited states -- due to charge fluctuations: from the
ground state to the electrode and to the excited state at QD.

The paper also brings some new insights in the
phase shift problem for the transmitted electrons. We find that the
electronic phase shows a specific dip at the excited state position,
in conditions of constant occupancy on QD. Similar features in
the phase evolution were reported in a recent quantum interference experiment performed by
Avinun-Kalish et al. \cite{Avinun}. This
experimental findings inspired  Rontani \cite{Rontani} to prove a
generalized Fridel sum rule. Our calculations are an explicit demonstration that
such dips in the phase evolution are due to electronic correlations in the
system, and a redistribution of charge. The phase measurement seems to be be a very
good detection method for such states, in a quantum interferometer device.

In the Kondo regime, there is a competition between the spin-spin
coupling with the magnetic impurity and the Kondo correlations with
the electrons from the leads. This competition was studied before in
a different system, namely a large isolated quantum dot, where it
was found that for the exchange parameter higher than a limit value,
the Kondo correlations are destroyed \cite{Murthy}. For the case of
a double-dot system, one can tune between a local singlet or a
double-Kondo situation by varying different parameters of the
system, like the interdot coupling \cite{Simon, Vavilov, hoff2006}.
Our model is different, but the conclusion is in agreement: we found
a limit value of the impurity-electron spin-spin correlator, that
inhibits the formation of the Kondo resonance. For the case when
both singlet and triplet levels are below the Fermi energy, the
spins are weakly correlated and we predict three Kondo peaks in the
density of states: one at $E_F$ and the other two at $E_F\pm J$,
corresponding to the excitation energy.

\acknowledgments{The work was supported by the project RTNNANO
contract No. MRTN-CT-2003-504574 and
by in part by the Ministry of Science and Higher Education (Poland).}
\vskip 1 cm

\appendix
\section{SPECTRUM OF THE ISOLATED DOT}

The isolated dot is described by the Hamiltonian (1), if we put
$t=0$. The energy spectrum is composed from the empty state,
the states with one and two electrons on the level $\epsilon_0$.
For single occupancy we have
\begin{eqnarray}
H|\uparrow\Uparrow\rangle&=&(\epsilon_0+\frac14J)|\uparrow\Uparrow\rangle~~(Triplet)\,,\nonumber\\
H|\downarrow\Downarrow\rangle&=&(\epsilon_0+\frac14J)|\downarrow\Downarrow\rangle~~(Triplet)\,,\nonumber\\
H\frac{1}{\sqrt{2}}(|\uparrow\Downarrow\rangle+|\downarrow\Uparrow\rangle)& =&(\epsilon_0+\frac14J)\frac{1}{\sqrt{2}}(|\uparrow\Downarrow\rangle+|\downarrow\Uparrow\rangle)~~(Triplet)\,,\\
H\frac{1}{\sqrt{2}}(|\uparrow\Downarrow\rangle-|\downarrow\Uparrow\rangle)& =&(\epsilon_0-\frac34J)\frac{1}{\sqrt{2}}(|\uparrow\Downarrow\rangle-|\downarrow\Uparrow\rangle)~~(Singlet)\,,\nonumber
\end{eqnarray}
where $|\uparrow\Uparrow\rangle$ is the state with the electron with the spin $\sigma=\uparrow$
at the state $\epsilon_0$ and the spin up at the impurity, etc.
For double occupancy, there is a two-fold degenerate state
\begin{eqnarray}
H|\uparrow\downarrow\Uparrow\rangle=(2\epsilon_0+U)|\uparrow\downarrow\Uparrow\rangle\,,\nonumber\\
H|\uparrow\downarrow\Downarrow\rangle=(2\epsilon_0+U)|\uparrow\downarrow\Downarrow\rangle\,.
\end{eqnarray}

For single occupancy, we see that there are two possible states. The
lowest in energy will be the ground state of the system (singlet for
positive $J$ and triplet for negative $J$), and the other will be
excited state. The peaks in transmittance correspond to the poles of the single-particle
Green function, which is found with the condition $\omega- E(\nu\,',n\pm1) + E(\nu,n)=0$
at $\omega=E_F$, where $E(\nu,n)$
is an energy state $\nu$ from the spectrum for QD with $n$ electrons (see e.g. \cite{Bruus}).
In our case,  for $E_F=0$, this is equivalent with imposing
the condition $E(\nu,n)=E(\nu\,',n\pm1)$, meaning that an energy level $\nu$ from the
spectrum with $n$ electrons becomes equal with an energy level $\nu\,'$ for $n\pm1$
electrons. This can be achieved by varying the diagonal energy
$\epsilon_0$ with an applied gate potential. For the adding of the first
electron we have the resonances at $\epsilon_0=3J/4$ and
$\epsilon_0=-J/4$; for the adding of the second electron we will
have two resonances at $\epsilon_0=-U+J/4$ and $\epsilon_0=-U-3J/4$.

\section{Discussion on the formation of the Kondo resonance}

In order to get a deeper insight on the formation of the Kondo resonances, one
can assume that, in the vicinity the singularity point $\omega=-a$, the
function $F_a$ (Eq.15) can be a dominating quantity in the Green function
formula, and an expansion in $1/F_a$ can be done. To be more clear, we
illustrate this idea first for the case $J=0$, when we have the single-impurity
Anderson Hamiltonian (within the approximations of Meir \cite{Meir91}, which is
similar to that of Lacroix's for high temperatures \cite{Lacroix}). We have
then

\begin{eqnarray}
\langle\langle c_{0\uparrow}|c_{0\uparrow}^\dagger\rangle\rangle^{J=0}=
\frac{1-\langle n_{0\downarrow}\rangle}
{\omega-\epsilon_0+i\Gamma +t^2 F_0}\,,\\
\frac1{\langle\langle
c_{0\uparrow}|c_{0\uparrow}^\dagger\rangle\rangle^{J=0}}=t^2\frac{1}{1-\langle
n_{0\downarrow}\rangle} F_0+...\;.
\end{eqnarray}
If the state $\epsilon_0$ is deep below the Fermi level $E_F$, so $\langle
n_{0\downarrow}\rangle\lesssim 0.5$, it results that the coefficient in front
of $F_0$ in the expression for $1/\langle\langle
c_{0\uparrow}|c_{0\uparrow}^\dagger\rangle\rangle$ is positive. It ensures the
existence of the Kondo peak. If we look at the denominator of the Green
function, we see that this is the condition for the "$\omega-\epsilon_0$" line
to intersect the logarithmical enhanced peak of $F_0$, and therefore, to give a
peak in the density of states near $E_F$ (see Fig.1 and Fig.2, and their interpretations,
in \cite{Lacroix}). Of course, in the simple case $J=0$, this is seen right away,
the reason why we expressed the expansion of $1/\langle\langle
c_{0\uparrow}|c_{0\uparrow}^\dagger\rangle\rangle$ is to compare the result
with the same expansion for $J\neq 0$, when the analytical formula of
$\langle\langle c_{0\uparrow}|c_{0\uparrow}^\dagger\rangle\rangle$ is much more
complex and not so transparent (but the expansion that we mentioned gives a
result easy to discuss).

In the case of finite $J$ we find
\begin{equation}
\frac1{\langle\langle
c_{0\uparrow}|c_{0\uparrow}^\dagger\rangle\rangle}=t^2\frac{1-\frac{1}{3}
\lambda\langle\vec{s}\cdot\vec{S}\rangle} {1-\langle
n_{0\downarrow}\rangle+\langle\vec{s}\cdot\vec{S}\rangle}F_0+0[\frac1{F_0}]...\;.
\end{equation}
The sign of the coefficient in front of $F_0$ depends on the sign of
the denominator $1-\langle
n_{0\downarrow}\rangle+\langle\vec{s}\cdot\vec{S}\rangle$, because the
numerator is always positive. In the Kondo regime, we have $\langle
n_{0\downarrow}\rangle\approx 0.5$ and the denominator changes its
sign for $\langle\vec{s}\cdot\vec{S}\rangle$, meaning that it is positive
for $\langle\vec{s}\cdot\vec{S}\rangle > -0.5$ and negative for
$\langle\vec{s}\cdot\vec{S}\rangle < -0.5$.

Three cases are presented in Fig.4: the case a) with
$\langle\vec{s}\cdot\vec{S}\rangle \simeq 0$, b)
$\langle\vec{s}\cdot\vec{S}\rangle \simeq 1/4$, and the case c) with
$\langle\vec{s}\cdot\vec{S}\rangle \simeq -3/4$. For the case a) the
coefficient in front of $F_0$ in the expansion (17) is positive, and
DOS shows the peak at $E_F$. In the case b), when the triplet is
below $E_F$ and singlet above $E_F$, the spin-spin average
$\langle\vec{s}\cdot\vec{S}\rangle$ is positive (with a maximum value
$1/4$), as the triplet state favors ferromagnetic coupling. For this
case again the coefficient in front of $F_0$ is positive. A
different situation is encountered in the case c), when singlet is
below $E_F$ and triplet is above $E_F$. The spin-spin average is
negative and, for large $J$, it can reach the minimal value -3/4. In
this case the coefficient of $F_0$ is negative, preventing the
formation of the Kondo resonance. According to our calculations, the
Kondo effect develops for $\langle\vec{s}\cdot\vec{S}\rangle\> > -0.5 $.


\begin{thebibliography}{99}
\bibitem{arctan} For angles greater than $\pi
/2$, we should compensate the artificial $\pi$ drop from a
definition of $\arctan$. One should compute ($\pi - \arctan$) for
the value of the phase.
\end{thebibliography}
\end{document}